\documentclass[aps, prl, amsmath, article]{revtex4}

\begin{document}

\title{Quantization of a string with attached mass}
\author{A. Lewis Licht   \footnote{licht@uic.edu}}
\affiliation{Dept. of Physics\\U. of Illinois at Chicago\\Chicago, Illinois 60607}

\begin{abstract}
We consider in the following the quantization of a simple model of a relativistic open string with a point mass attached at one end.  The normal modes are derived and used to construct expressions for the position-position and position-conjugate commutators.  Light cone gauge is used to find the mass squared operator.  The singular part of the operator product expansion is derived.
\end{abstract}

\maketitle

\section{1.   Introduction}\label{S : intro}

We consider in the following a relativistic bosonic open string with a point mass attached at one end.  In a following paper ~\cite{all-1} we will discuss a string with point masses attached at each end.  This might be considered a model for a meson, as also discussed in the Lund model~\cite{Lund-1}, but incorporating some of the techniques of modern string theory.~\cite{GSW-1}~\cite{Polch-1}~\cite{BBS-1}

In Section 2 we derive the oscillation modes and also an equation giving the allowed frequencies.  The boundary condition due to the attached mass makes the standard canonical commutation relations invalid.  The system could possibly be quantized using Dirac brackets~\cite{Dirac-1}, but in Section 3 we find it more convenient to quantize the mode amplitudes.~\cite{ModeQ-1} The resulting expressions for the position-velocity and position-conjugate commutators are given a simplified form in Section 4.  In Section 5 light cone gauge is used to find the expression for the mass squared operator.  In Section 6 the singular part of the operator product expansion is derived. 

\section{2.  The Oscillation Modes}

An open string has the action

\begin{equation}
S = \frac{{T_0 }}
{2}\int {d\tau } \int_0^\pi  {d\sigma } \left( {\dot X^2 \left( {\tau ,\sigma } \right) - X'^2 \left( {\tau ,\sigma } \right)} \right)
\end{equation}

Here the parameter $T_0 $  has the role of a mass per unit length when it multiplies the time derivative term, and also is the coefficient of tension when it multiplies the spatial derivative term.  We introduce a point mass at the  $\sigma  = \pi $ end by increasing there the mass per unit length   by a delta function distribution of strength  $m_0$.  The string with an attached point mass at one end then has the action:

\begin{equation}\label{E: mSact}
S = \frac{{m_0 }}
{2}\int {d\tau } \dot X^2 \left( {\tau ,\pi } \right) + \frac{{T_0 }}
{2}\int {d\tau } \int_0^\pi  {d\sigma } \left( {\dot X^2 \left( {\tau ,\sigma } \right) - X'^2 \left( {\tau ,\sigma } \right)} \right)
\end{equation}

It obeys the boundary conditions
\begin{equation}\label{E: BC}
\begin{gathered}
  m_0 \ddot X^\mu  \left( {\tau ,\pi } \right) =  - T_0 X'^\mu  \left( {\tau ,\pi } \right) \\ 
  X'^\mu  \left( {\tau ,0} \right) = 0 \\ 
\end{gathered} 
\end{equation}

and the equation of motion
\begin{equation}
\ddot X^\mu  \left( {\tau ,\sigma } \right) - X''^\mu  \left( {\tau ,\sigma } \right) = 0
\end{equation}

There is a linear solution:
\begin{equation}\label{E: linear}
X_0^\mu   = x^\mu   + b^\mu  \tau 
\end{equation}

and also oscillating solutions:
\begin{equation}\label{E: XtoB}
X_\omega ^\mu  \left( {\tau ,\sigma } \right) = B_\omega ^\mu  \left( \tau  \right)\cos \left( {\omega \sigma } \right)
\end{equation}

where
\begin{equation}\label{E: BtoA}
B_\omega ^\mu  \left( \tau  \right) = A_{\omega  + }^\mu  e^{ - i\omega \tau }  + A_{\omega  - }^\mu  e^{ + i\omega \tau } 
\end{equation}

and the frequencies $\omega $  must satisfy
\begin{equation}\label{E: Ttom}
T_0 \sin \left( {\omega \pi } \right) =  - m_0 \omega \cos \left( {\omega \pi } \right)
\end{equation}
as a consequence of Eq.~(\ref{E: BC} a).

\section{3.   Quantization}

We write
\begin{equation}
X^\mu  \left( {\tau ,\sigma } \right) = X_0^\mu  \left( \tau  \right) + \sum\limits_{\omega  > 0} {} X_\omega ^\mu  \left( {\tau ,\sigma } \right)
\end{equation}
and substitute this expression in Eq.~(\ref{E: mSact})
 The terms in the Lagrangian  involving $X_0^\mu  $  are 
\begin{equation}
\begin{gathered}
  \frac{{m_0 }}
{2}\left( {\dot X_0 } \right)^2  + \frac{{T_0 \pi }}
{2}\left( {\dot X_0 } \right)^2  + m_0 \dot X_0^\mu  \sum\limits_{\omega  > 0} {} \dot B_{\omega \mu } \cos \left( {\omega \pi } \right) + T_0 \dot X_0^\mu  \sum\limits_{\omega  > 0} {} \dot B_{\omega \mu } \frac{{\sin \left( {\omega \pi } \right)}}
{\omega } \hfill \\
  \quad \quad \quad \quad \quad \quad \quad  = \frac{{m_0 }}
{2}\left( {\dot X_0 } \right)^2  + \frac{{T_0 \pi }}
{2}\left( {\dot X_0 } \right)^2  \hfill \\ 
\end{gathered} 
\end{equation}
where we have used Eq.~(\ref{E: Ttom})

The other time derivative terms are
\begin{equation}
TD = \sum\limits_{\omega ,\omega ' > 0} {} \dot B_\omega   \cdot \dot B_{\omega '} \left[ {\frac{{m_0 }}
{2}\cos \left( {\omega \pi } \right)\cos \left( {\omega '\pi } \right) + \frac{{T_0 }}
{2}\int_0^\pi  {d\sigma } \cos \left( {\omega \sigma } \right)\cos \left( {\omega '\sigma } \right)} \right]
\end{equation}
Now if $\omega  \ne \omega '$  then
\begin{equation}
\begin{gathered}
  \frac{{m_0 }}
{2}\cos \left( {\omega \pi } \right)\cos \left( {\omega '\pi } \right) + \frac{{T_0 }}
{2}\int_0^\pi  {d\sigma } \cos \left( {\omega \sigma } \right)\cos \left( {\omega '\sigma } \right) \hfill \\
  \quad \quad  = \frac{{m_0 }}
{2}\cos \left( {\omega \pi } \right)\cos \left( {\omega '\pi } \right) + \frac{{T_0 }}
{{2\left( {\omega ^2  - \omega '^2 } \right)}}\left[ {\omega \sin \left( {\omega \pi } \right)\cos \left( {\omega '\pi } \right) - \omega '\sin \left( {\omega '\pi } \right)\cos \left( {\omega \pi } \right)} \right] \hfill \\
  \quad \quad  = 0 \hfill \\ 
\end{gathered} 
\end{equation}
where we have used Eq.~(\ref{E: Ttom})

When $\omega  = \omega '$  then
\begin{equation}
\begin{gathered}
  \frac{{m_0 }}
{2}\cos ^2 \left( {\omega \pi } \right) + \frac{{T_0 }}
{2}\int_0^\pi  {d\sigma } \cos ^2 \left( {\omega \sigma } \right) \hfill \\
  \quad \quad  = \frac{{m_0 }}
{2}\cos ^2 \left( {\omega \pi } \right) + \frac{{T_0 }}
{2}\left[ {\frac{\pi }
{2} + \frac{{\sin \left( {\omega \pi } \right)\cos \left( {\omega \pi } \right)}}
{{2\omega }}} \right] \hfill \\
  \quad \quad  = \frac{1}
{2}\left[ {\frac{{m_0 }}
{2}\cos ^2 \left( {\omega \pi } \right) + \frac{{T_0 \pi }}
{2}} \right] \hfill \\ 
\end{gathered} 
\end{equation}
Again using Eq.~(\ref{E: Ttom}).

The spatial derivative terms are
\begin{equation}
SD =  - \frac{{T_0 }}
{2}\sum\limits_{\omega ,\omega ' > 0}^{} {} B_\omega   \cdot B_{\omega '} \int_0^\pi  {d\sigma } \omega \omega '\sin \left( {\omega \sigma } \right)\sin \left( {\omega '\sigma } \right)
\end{equation}
Now if $\omega  \ne \omega '$  then
\begin{equation}
\begin{gathered}
  T_0 \int_0^\pi  {d\sigma } \sin \left( {\omega \sigma } \right)\sin \left( {\omega '\sigma } \right) = \frac{{T_0 }}
{2}\int_0^\pi  {} \left[ {\cos \left( {\left( {\omega  - \omega '} \right)\pi } \right) - \cos \left( {\left( {\omega  + \omega '} \right)\pi } \right)} \right] \hfill \\
  \quad  = \frac{{T_0 }}
{2}\left[ {\frac{{\sin \left( {\left( {\omega  - \omega '} \right)\pi } \right)}}
{{\omega  - \omega '}} - \frac{{\sin \left( {\left( {\omega  + \omega '} \right)\pi } \right)}}
{{\omega  + \omega '}}} \right] \hfill \\
  \quad  = \frac{{T_0 }}
{2}\left[ {\frac{{\sin \left( {\omega \pi } \right)\cos \left( {\omega '\pi } \right) - \sin \left( {\omega '\pi } \right)\cos \left( {\omega \pi } \right)}}
{{\omega  - \omega '}} - \frac{{\sin \left( {\omega \pi } \right)\cos \left( {\omega '\pi } \right) + \sin \left( {\omega '\pi } \right)\cos \left( {\omega \pi } \right)}}
{{\omega  + \omega '}}} \right] \hfill \\
  \quad  = 0 \hfill \\ 
\end{gathered} 
\end{equation}

by Eq.~(\ref{E: Ttom})

When  $\omega  = \omega '$ then a typical factor is .
\begin{equation}
\begin{gathered}
   - \frac{{T_0 }}
{2}\omega ^2 \int_0^\pi  {d\sigma } \sin ^2 \left( {\omega \sigma } \right) =  - \frac{{T_0 }}
{4}\omega ^2 \int_0^\pi  {d\sigma } \left( {1 - \cos \left( {2\omega \sigma } \right)} \right) \\ 
   =  - \frac{{T_0 }}
{4}\omega ^2 \left( {\pi  - \frac{{\sin \left( {\omega \pi } \right)\cos \left( {\omega \pi } \right)}}
{\omega }} \right) \\ 
   =  - \frac{{T_0 }}
{4}\omega ^2 \pi  - \frac{{m_0 }}
{4}\omega ^2 \cos ^2 \left( {\omega \pi } \right) \\ 
\end{gathered} 
\end{equation}
The Lagrangian can now be written as
\begin{equation}
L = \frac{1}
{2}\left( {m_0  + T_0 \pi } \right)\left( {\dot X_0 } \right)^2  + \frac{1}
{4}\sum\limits_{\omega  > 0} {} \left( {m_0 \cos ^2 \left( {\omega \pi } \right) + T_0 \pi } \right)\left[ {\left( {\dot B_\omega  } \right)^2  - \omega ^2 \left( {B_\omega  } \right)^2 } \right]
\end{equation}

From this we can find the canonical conjugates:
\begin{equation}\label{E: canconj}
\begin{gathered}
  \Pi _{\mu 0}  = \frac{{\partial L}}
{{\partial \dot X_0^\mu  }} = \left( {m_0  + T_0 \pi } \right)\dot X_{\mu 0}  \hfill \\
  \Pi _{\mu \omega }  = \frac{{\partial L}}
{{\partial \dot X_\omega ^\mu  }} = \frac{1}
{2}\left( {m_0 \cos ^2 \left( {\omega \pi } \right) + T_0 \pi } \right)\dot B_{\mu \omega }  \hfill \\ 
\end{gathered} 
\end{equation}
The canonical commutation rules are
\begin{equation}
\begin{gathered}
  \left[ {X_0^\mu  ,\Pi _{\nu 0} } \right] = i\delta _\nu ^\mu   \\ 
  \left[ {X_\omega ^\mu  ,\Pi _{\nu \omega '} } \right] = i\delta _\nu ^\mu  \delta _{\omega \omega '}  \\ 
\end{gathered} 
\end{equation}
Which for the zero mode implies that
\begin{equation}
\left[ {X_0^\mu  ,\dot X_0^\nu  } \right] = i\frac{{\eta ^{\mu \nu } }}
{{m_0  + T_0 \pi }}
\end{equation}
and therefore with Eq.~(\ref{E: linear}) we can identify
\begin{equation}
b^\nu   = \frac{{p^\nu  }}
{{m_0  + T_0 \pi }}
\end{equation}
where $p^\nu  $  is the total momentum operator, with $\left[ {x^\mu  ,p^\nu  } \right] = i\eta ^{\mu \nu } $.
With  $Q\left( \omega  \right) = m_0 \cos ^2 \left( {\omega \pi } \right) + T_0 \pi $
 , the non-zero modes satisfy  
\begin{equation}
\left[ {B_\omega ^\mu  \left( \tau  \right),\dot B_{\omega '}^\nu  \left( \tau  \right)} \right] = 2i\frac{{\eta ^{\mu \nu } }}
{{Q\left( \omega  \right)}}\delta _{\omega \omega '} 
\end{equation}
Using Eq.~(\ref{E: BtoA}), this leads to 
\begin{equation}
\begin{gathered}
  \left[ {A_{\omega  + }^\mu  ,A_{\omega ' + }^\nu  } \right] = \left[ {A_{\omega  - }^\mu  ,A_{\omega ' - }^\nu  } \right] = 0 \hfill \\
  \left[ {A_{\omega  + }^\mu  ,A_{\omega ' - }^\nu  } \right] = \frac{{\eta ^{\mu \nu } }}
{{\omega Q\left( \omega  \right)}}\delta _{\omega \omega '}  \hfill \\ 
\end{gathered} 
\end{equation}
Define, for $\omega  > 0$
\begin{equation}\label{E: alphap}
\alpha _\omega ^\mu   =  - i\omega \sqrt {Q\left( \omega  \right)} A_{\omega  + }^\mu  
\end{equation}
and for $\omega  < 0$
\begin{equation}\label{E: alpham}
\alpha _\omega ^\mu   =  - i\omega \sqrt {Q\left( \omega  \right)} A_{\left| \omega  \right| - }^\mu  
\end{equation}
Then we get the conventional commutator:
\begin{equation}
\left[ {\alpha _\omega ^\mu  ,\alpha _{\omega '}^\nu  } \right] = \omega \eta ^{\mu \nu } \delta _{\omega  + \omega ',0} 
\end{equation}
And we can now write the coordinate operator as 
\begin{equation}\label{E: CordOp}
X^\mu  \left( {\tau ,\sigma } \right) = x^\mu   + \frac{{p^\mu  \tau }}
{{m_0  + T_0 \pi }} + i\sum\limits_{\omega  \ne 0} {} \frac{{\alpha _\omega ^\mu  }}
{{\omega \sqrt {m_0 \cos ^2 \left( {\omega \pi } \right) + T_0 \pi } }}e^{ - i\omega \tau } \cos \left( {\omega \sigma } \right)
\end{equation}
The equal time position-velocity commutator is
\begin{equation}
\begin{gathered}
  \left[ {X^\mu  \left( {\tau ,\sigma } \right),\dot X^\nu  \left( {\tau ,\sigma '} \right)} \right] = i\eta ^{\mu \nu } \left[ {\frac{1}
{{m_0  + T_0 \pi }} + 2\sum\limits_{\omega  > 0} {} \frac{{\cos \left( {\omega \sigma } \right)\cos \left( {\omega \sigma '} \right)}}
{{m_0 \cos ^2 \left( {\omega \pi } \right) + T_0 \pi }}} \right] \\ 
   = i\eta ^{\mu \nu } D\left( {\sigma ,\sigma '} \right) \\ 
\end{gathered} 
\end{equation}
We will examine this in detail in the following.

\section{4.   The Commutator}

The action of Eq.~(\ref{E: mSact}) gives for $\Pi _\mu  \left( {\tau ,\sigma } \right)$ ,  the operator canonical conjugate to $X^\mu  \left( {\tau ,\sigma } \right)$  ,
\begin{equation}
\begin{gathered}
  \Pi _\mu  \left( {\tau ,\sigma } \right) = \frac{{\delta S}}
{{\delta \dot X_\mu  \left( {\tau ,\sigma } \right)}} \\ 
   = m_0 \dot X_\mu  \left( {\tau ,\pi } \right)\delta \left( {\sigma  - \pi } \right) + T_0 \dot X_\mu  \left( {\tau ,\sigma } \right)|_{\sigma  < \pi }  \\ 
   = m_0 \left[ {\frac{{p_\mu  }}
{{m_0  + T_0 \pi }} + \sum\limits_{\omega  \ne 0} {} \frac{{\alpha _{\mu \omega } }}
{{\sqrt {m_0 \cos ^2 \left( {\omega \pi } \right) + T_0 \pi } }}e^{ - i\omega \tau } \cos \left( {\omega \pi } \right)} \right]\delta \left( {\sigma  - \pi } \right) \\ 
  \quad  + T_0 \left[ {\frac{{p_\mu  }}
{{m_0  + T_0 \pi }} + \sum\limits_{\omega  \ne 0} {} \frac{{\alpha _{\mu \omega } }}
{{\sqrt {m_0 \cos ^2 \left( {\omega \pi } \right) + T_0 \pi } }}e^{ - i\omega \tau } \cos \left( {\omega \sigma } \right)} \right] \\ 
\end{gathered} 
\end{equation}
Then the position-conjugate commutator is therefore	
\begin{equation}
\left[ {X^\mu  \left( {\tau ,\sigma } \right),\Pi ^\nu  \left( {\tau ,\sigma '} \right)} \right] = i\eta ^{\mu \nu } \left[ {m_0 D\left( {\sigma ,\pi } \right)\delta \left( {\sigma ' - \pi } \right) + T_0 D\left( {\sigma ,\sigma '} \right)|_{\sigma ' < \pi } } \right]_{\sigma  < \pi } 
\end{equation}
The commutator function  $D\left( {\sigma ,\sigma '} \right)$ can be written as
\begin{equation}
D\left( {\sigma ,\sigma '} \right) = \sum\limits_{n =  - \infty }^{ + \infty } {} \frac{{\cos \left( {\omega _n \sigma } \right)\cos \left( {\omega _n \sigma '} \right)}}
{{m_0 \cos ^2 \left( {\omega _n \pi } \right) + T_0 \pi }}
\end{equation}
where we have labeled the frequencies  $\omega _n $ given by Eg.~(\ref{E: Ttom}) in the order of their magnitude, with $\omega _n  < \omega _{n + 1} $  and $\omega _0  = 0$  .  The function
\begin{equation}
f\left( \omega  \right) = \frac{1}
{{\frac{{m_0 \omega }}
{{T_0 }} + \tan \left( {\omega \pi } \right)}}
\end{equation}
has poles at each $\omega _n $  with residue
\begin{equation}
\begin{gathered}
  Lim_{\omega  \to \omega _n } \frac{{\omega  - \omega _n }}
{{\frac{{m_0 \omega }}
{{T_0 }} + \tan \left( {\omega \pi } \right)}} = \frac{1}
{{\frac{d}
{{d\omega }}\left[ {\frac{{m_0 \omega }}
{{T_0 }} + \tan \left( {\omega \pi } \right)} \right]_{\omega _n } }} \\ 
   = \frac{{T_0 \cos ^2 \left( {\omega _n \pi } \right)}}
{{m_0 \cos ^2 \left( {\omega _n \pi } \right) + T_0 \pi }} \\ 
\end{gathered} 
\end{equation}
Therefore, let  $C_n $ denote a very small circular path in the complex plane around each real $\omega _n $  , then we can write
\begin{equation}\label{E: Dss}
D\left( {\sigma ,\sigma '} \right) = \sum\limits_{n =  - \infty }^{ + \infty } {} \oint_{C_n } {\frac{{d\omega }}
{{2\pi i}}} \frac{{\cos \left( {\omega \sigma } \right)\cos \left( {\omega \sigma '} \right)}}
{{T_0 \cos ^2 \left( {\omega \pi } \right)\left[ {\frac{{m_0 \omega }}
{{T_0 }} + \tan \left( {\omega \pi } \right)} \right]}}
\end{equation}
Next we expand the circles  $C_n $ to become two lines parallel to the real axis, one above and one below the real axis, plus a series of new, clockwise paths $C'_k $  around each of the zeros of  $\cos \left( {\omega \pi } \right)$.   The parallel lines may be taken to infinity, where each integrand decreases as  $\exp \left[ { - \left| {\operatorname{Im} \left( \omega  \right)} \right|\left( {2\pi  - \sigma  - \sigma '} \right)} \right]$ .  There are three possibilities:  (1) Both  $\sigma $ and $\sigma '$  are each less than $\pi $ .  (2)  One, say $\sigma  < \pi $  , but  $\sigma ' = \pi $ .  (3)  $\sigma  = \sigma ' = \pi $  .

(1)  The parallel lines contribute nothing.  The zeros occur at  $\omega _k  = \frac{{2k + 1}}
{2}$  for integers k, and with  $\omega  = \omega _k  + \varepsilon $  near such a point, we have

\begin{equation}
\begin{gathered}
  T_0 \cos ^2 \left( {\omega \pi } \right)\left[ {\frac{{m_0 \omega }}
{{T_0 }} + \tan \left( {\omega \pi } \right)} \right] = \cos \left( {\omega \pi } \right)\left[ {m_0 \omega \cos \left( {\omega \pi } \right) + T_0 \sin \left( {\omega \pi } \right)} \right] \\ 
   =  - \left( { - 1} \right)^k \varepsilon \pi \left[ {m_0 O\left( \varepsilon  \right) + T_0 \left( { - 1} \right)^k } \right] \\ 
   =  - \varepsilon \pi T_0  \\ 
\end{gathered} 
\end{equation}

The minus sign converts the clockwise paths to counterclockwise, giving us
\begin{equation}
\begin{gathered}
  D\left( {\sigma ,\sigma '} \right) = \frac{2}
{{T_0 \pi }}\sum\limits_{k = 0} {} \cos \left( {\frac{{2k + 1}}
{2}\sigma } \right)\cos \left( {\frac{{2k + 1}}
{2}\sigma '} \right) \\ 
   = \frac{1}
{{T_0 }}\delta \left( {\sigma  - \sigma '} \right) \\ 
\end{gathered} 
\end{equation}

since the functions $\sqrt {\frac{2}{\pi }} \cos \left( {\frac{{2k + 1}}{2}\sigma } \right)$  are normal, orthogonal and complete on the interval $0 < \sigma  < \pi $ .

(2)  The parallel lines still contribute nothing, but now the integrand is 

\begin{equation}
\begin{gathered}
  D\left( {\sigma ,\pi } \right) = \sum\limits_{n =  - \infty }^{ + \infty } {} \oint_{C_n } {\frac{{d\omega }}
{{2\pi i}}} \frac{{\cos \left( {\omega \sigma } \right)\cos \left( {\omega \pi } \right)}}
{{T_0 \cos ^2 \left( {\omega \pi } \right)\left[ {\frac{{m_0 \omega }}
{{T_0 }} + \tan \left( {\omega \pi } \right)} \right]}} \\ 
   = \sum\limits_{n =  - \infty }^{ + \infty } {} \oint_{C_n } {\frac{{d\omega }}
{{2\pi i}}} \frac{{\cos \left( {\omega \sigma } \right)}}
{{T_0 \cos \left( {\omega \pi } \right)\left[ {\frac{{m_0 \omega }}
{{T_0 }} + \tan \left( {\omega \pi } \right)} \right]}} \\ 
\end{gathered} 
\end{equation}
However the denominator in the integrand near the zeros of  $\cos \left( {\omega \pi } \right)$
 becomes
\begin{equation}
T_0 \cos \left( {\omega \pi } \right)\left[ {\frac{{m_0 \omega }}
{{T_0 }} + \tan \left( {\omega \pi } \right)} \right] \to m_0 O\left( \varepsilon  \right) + T_0 \left( { - 1} \right)^k 
\end{equation}
There are therefore no poles other than those at  $\omega _n $ and thus
\begin{equation}
D\left( {\sigma ,\pi } \right) = D\left( {\pi ,\sigma } \right) = 0
\end{equation}

(3)  In this case there are no extra poles in the integrand, but the parallel lines do contribute.  The commutator function is now
\begin{equation}
\begin{gathered}
  D\left( {\pi ,\pi } \right) = \sum\limits_{n =  - \infty }^{ + \infty } {} \frac{{\cos ^2 \left( {\omega _n \pi } \right)}}
{{m_0 \cos ^2 \left( {\omega _n \pi } \right) + T_0 \pi }} \\ 
   = \sum\limits_{n =  - \infty }^{ + \infty } {} \oint_{C_n } {\frac{{d\omega }}
{{2\pi i}}} \frac{1}
{{T_0 \left[ {\frac{{m_0 \omega }}
{{T_0 }} + \tan \left( {\omega \pi } \right)} \right]}} \\ 
\end{gathered} 
\end{equation}
We can now expand the integrations over the $C_n $  into one circle of radius R, and take the limit as  $R \to \infty $ .  Then 
\begin{equation}
\begin{gathered}
  D\left( {\pi ,\pi } \right) = \frac{1}
{{2\pi im_0 }}\oint_R {\frac{{d\omega }}
{\omega }}  \\ 
   = \frac{1}
{{m_0 }} \\ 
\end{gathered} 
\end{equation}
Summarising, we can now write the commutator function as
\begin{equation}
D\left( {\sigma ,\sigma '} \right) = \frac{1}
{{m_0 }}\delta _{\sigma ,\pi } \delta _{\sigma ',\pi }  + \frac{1}
{{T_0 }}\delta _\pi  \left( {\sigma ,\sigma '} \right)
\end{equation}

where $\delta _{\sigma ,\pi } $  is an ordinary Kronecker delta, and  $\delta _\pi  \left( {\sigma ,\sigma '} \right) = \delta \left( {\sigma  - \sigma '} \right)$ and is non zero, so long as neither of its arguments equals $\pi $ .

The position-conjugate commutator is now

\begin{equation}
\left[ {X^\mu  \left( {\tau ,\sigma } \right),\Pi ^\nu  \left( {\tau ,\sigma '} \right)} \right] = i\eta ^{\mu \nu } \left[ {\delta _{\sigma ,\pi } \delta \left( {\sigma ' - \pi } \right) + \delta _\pi  \left( {\sigma ,\sigma '} \right)} \right]
\end{equation}

\section{5.   The Mass Squared Operator}

We determine the possible physical states by going into the light cone frame.  In this frame $X^ +   = \frac{{X^0  + X^1 }}{{\sqrt 2 }}$  ,  $X^ -   = \frac{{X^0  - X^1 }}{{\sqrt 2 }}$  , and the remaining coordinates are  $X^I $ , for $I = 2,3, \cdots D - 2$  .  We then fix the gauge by setting
\begin{equation}
X^ +   = \tau
\end{equation}
From Eq.~(\ref{E: CordOp}) we see that this requires
\begin{equation}
\begin{gathered}\label{E: pluscord}
  p^ +   = m_0  + T_0 \pi  \\ 
  x^ +   = 0 \\ 
  \alpha _\omega ^ +   = 0 \\ 
\end{gathered} 
\end{equation}
(for $\omega  \ne 0$ ), and the Lagrangian becomes
\begin{equation}
\begin{gathered}
  L = \frac{{m_0 }}
{2}\left[ { - 2\dot X^ -  \left( {\tau ,\pi } \right) + \left( {\dot X^I \left( {\tau ,\pi } \right)} \right)^2 } \right] \hfill \\
  \quad \quad \quad  + \frac{{T_0 }}
{2}\int_0^\pi  {d\sigma } \left[ { - 2\dot X^ -  \left( {\tau ,\sigma } \right) + \left( {\dot X^I \left( {\tau ,\sigma } \right)} \right)^2  - \left( {X'^I \left( {\tau ,\sigma } \right)} \right)^2 } \right] \hfill \\ 
\end{gathered} 
\end{equation}

From Eq.~(\ref{E: CordOp}) we also see that the oscillatory parts of  $X^ -  \left( {\tau ,\sigma } \right)$  appear in this Lagrangian as

\begin{equation}
\begin{gathered}
   - m_0 \frac{{\alpha _\omega ^ -  }}
{{\sqrt {Q\left( \omega  \right)} }}e^{ - i\omega \tau } \cos \left( {\omega \pi } \right) - T_0 \int_0^\pi  {d\sigma } \frac{{\alpha _\omega ^ -  }}
{{\sqrt {Q\left( \omega  \right)} }}e^{ - i\omega \tau } \cos \left( {\omega \sigma } \right) \\ 
   =  - \frac{{\alpha _\omega ^ -  }}
{{\sqrt {Q\left( \omega  \right)} }}e^{ - i\omega \tau } \left( {m_0 \cos \left( {\omega \pi } \right) + \frac{{T_0 }}
{\omega }\sin \left( {\omega \pi } \right)} \right) = 0 \\ 
\end{gathered} 
\end{equation}

where we have used Eq.~(\ref{E: Ttom}).  The oscillatory part of  $X^ -  $ thus does not appear in the Lagrangian and may be set to 0.  The Lagrangian becomes
\begin{equation}
\begin{gathered}
  L =  - \left( {m_0  + T_0 \pi } \right)\dot X_0^ -   + \frac{{m_0 }}
{2}\left( {\dot X^I \left( {\tau ,\pi } \right)} \right)^2  \hfill \\
  \quad \quad \quad  + \frac{{T_0 }}
{2}\int_0^\pi  {d\sigma } \left[ {\left( {\dot X^I \left( {\tau ,\sigma } \right)} \right)^2  - \left( {X'^I \left( {\tau ,\sigma } \right)} \right)^2 } \right] \hfill \\ 
\end{gathered} 
\end{equation}

This gives us 
\begin{equation}
p_ -   = \frac{{\partial L}}
{{\partial \dot X^ -  }} =  - \left( {m_0  + T_0 \pi } \right) =  - p^ +  
\end{equation}
which is consistent with Eq.~(\ref{E: pluscord}).  With the  components expanded as in Eq.~(\ref{E: XtoB}) the Lagrangian becomes

\begin{equation}
L = \left( {m_0  + T_0 \pi } \right)\left( { - \dot X_0^ -   + \frac{{\left( {\dot X_0^I } \right)^2 }}
{2}} \right) + \frac{1}
{4}\sum\limits_{\omega  > 0} {} Q\left( \omega  \right)\left[ {\left( {\dot B_\omega ^I } \right)^2  - \omega ^2 \left( {B_\omega ^I } \right)^2 } \right]
\end{equation}

The canonical momenta of Eq.~(\ref{E: canconj}) are now
\begin{equation}
\begin{gathered}\label{E: newcanm}
  p_0^I  = \frac{{\partial L}}
{{\partial \dot X_0^I }} = Q\left( 0 \right)\dot X_0^I  \hfill \\
  p_\omega ^I  = \frac{{\partial L}}
{{\partial \dot B_\omega ^I }} = \frac{{Q\left( \omega  \right)}}
{2}\dot B_\omega ^I  \hfill \\ 
\end{gathered} 
\end{equation}
From this we obtain the Hamiltonian,
\begin{equation}
\begin{gathered}\label{E: Ham}
  H = p_ -  \dot X_0^ -   + p_0^I \dot X_0^I  + \sum\limits_{\omega  > 0} {} p_\omega ^I \dot B_\omega ^I  - L \\ 
   = \frac{{\left( {p_0^I } \right)^2 }}
{{2Q\left( 0 \right)}} + \sum\limits_{\omega  > 0} {} \left[ {\frac{{\left( {p_\omega ^I } \right)^2 }}
{{Q\left( \omega  \right)}} + \frac{{\omega ^2 Q\left( \omega  \right)\left( {B_\omega ^I } \right)^2 }}
{4}} \right] \\ 
\end{gathered} 
\end{equation}
We identify $H = p^ -   =  - p_ +  $  as it is the operator that changes $\tau $ . (\cite{Polch-1}, vol I, p 17).  Then using Eq.~(\ref{E: newcanm}) and with $Q\left( 0 \right) = m_0  + T_0 \pi  = p^ +  $ ,  Eq.~(\ref{E: Ham}) becomes
\begin{equation}
\begin{gathered}
  M^2  = 2p^ +  p^ -   - \left( {p_0^I } \right)^2  \\ 
   = \frac{{\left( {m_0  + T_0 \pi } \right)}}
{2}\sum\limits_{\omega  > 0} {} Q\left( \omega  \right)\left[ {\left( {\dot B_\omega ^I } \right)^2  + \omega ^2 \left( {B_\omega ^I } \right)^2 } \right] \\ 
\end{gathered} 
\end{equation}
From Eqs.~(\ref{E: BtoA}) and (\ref{E: alphap}, \ref{E: alpham}) we can write
\begin{equation}
B_\omega ^I  = \frac{i}
{{\omega \sqrt {Q\left( \omega  \right)} }}\left[ {\alpha _\omega ^I e^{ - i\omega \tau }  - \alpha _{ - \omega }^I e^{ + i\omega \tau } } \right]
\end{equation}
Then
\begin{equation}\label{E: m2}
M^2  = 2Q\left( 0 \right)\sum\limits_{\omega  > 0} {} \alpha _{ - \omega }^I \alpha _\omega ^I  + \left( {D - 2} \right)A
\end{equation}
where
\begin{equation}
A = Q\left( 0 \right)\sum\limits_{\omega  > 0} {} \omega 
\end{equation}
For  $m_0  \gg T_0 $ it is convenient to write the solutions to Eq.~(\ref{E: Ttom}) as
\begin{equation}
\omega _k  = \frac{{2k + 1}}
{2} + \varepsilon _k 
\end{equation}
where $\varepsilon _k $  satisfies
\begin{equation}
\tan \left( {\varepsilon _k \pi } \right) = \frac{{T_0 }}
{{m_0 \left( {\frac{{2k + 1}}
{2} + \varepsilon _k } \right)}}
\end{equation}
For  $m_0  \gg T_0 $ this becomes
\begin{equation}
\varepsilon _k  = \frac{{2T_0 }}
{{\pi m_0 \left( {2k + 1} \right)}} + O\left( {\left( {\frac{{T_0 }}
{{m_0 }}} \right)^2 } \right)
\end{equation}
and Eq. ~(\ref{E: m2}) becomes
\begin{equation}
M^2  = 2\left( {m_0  + T_0 \pi } \right)\left[ {\sum\limits_{\omega  > 0} {} \alpha _{ - \omega }^I \alpha _\omega ^I  + \frac{{D - 2}}
{2}\sum\limits_{k = 0} {} \left( {\frac{{2k + 1}}
{2} + \frac{{2T_0 }}
{{\pi m_0 \left( {2k + 1} \right)}}} \right)} \right]
\end{equation}
We regularize the third sum by considering
\begin{equation}
\begin{gathered}
  \sum\limits_{k = 0}^\infty  {} \frac{2}
{{2k + 1}}e^{ - k\varepsilon }  = \ln \left( {1 + e^{ - \varepsilon } } \right) - \ln \left( {1 - e^{ - \varepsilon } } \right) \\ 
   = \ln \left( 2 \right) - 2\ln \left( \varepsilon  \right) - \frac{{\varepsilon ^2 }}
{6} +  \cdots  \\ 
\end{gathered} 
\end{equation}
Taking the limit as  $\varepsilon  \to 0$ and discarding the divergent part, this becomes simply ln~(2).
We write the second sum as
\begin{equation}
 - \partial _\varepsilon  \sum\limits_{k = 0} {} e^{ - \frac{{2k + 1}}
{2}\varepsilon }  = \frac{1}
{{\varepsilon ^2 }} + \frac{1}
{{24}} + O\left( \varepsilon  \right)
\end{equation}
Discarding the divergent term and taking the limit as $\varepsilon  \to 0$  we get
\begin{equation}
M^2  = 2\left( {m_0  + T_0 \pi } \right)\left[ {\sum\limits_{\omega  > 0} {} \alpha _{ - \omega }^I \alpha _\omega ^I  + \frac{{D - 2}}
{2}\left( {\frac{1}
{{24}} + \frac{{T_0 }}
{{\pi m_0 }}\ln \left( 2 \right) + O\left( {\left( {\frac{{T_0 }}
{{m_0 }}} \right)^2 } \right)} \right)} \right]
\end{equation}

Unlike the situation for the open string without attached mass, the ground state here has positive mass squared.  
If  $m_0 $ is not much greater than  $T_0 $ the situation is more complicated.  If it is much less than   $T_0 $ it can be shown that the frequencies  $\omega _n $ are slightly less than the integers n, so long as n is not too large, and eventually, for large n, they become as above, slightly greater than the half odd integers.

\section{6.  The Operator Product Expansion}

We map the string world sheet coordinates  $\left( {\tau ,\sigma } \right)$ into the complex plane by setting

\begin{equation}
\begin{gathered}
  \tau  =  - i\sigma ^0 ,\quad \quad \sigma  = \sigma ^1  \\ 
  z = e^{\sigma ^0  + i\sigma ^1 }  = e^{i\left( {\tau  + \sigma } \right)}  \\ 
  \bar z = e^{\sigma ^0  - i\sigma ^1 }  = e^{i\left( {\tau  - \sigma } \right)}  \\ 
\end{gathered} 
\end{equation}

Then the operator expansion of Eq.~(\ref{E: CordOp}) becomes
\begin{equation}
X^\mu  \left( {\tau ,\sigma } \right) \to X^\mu  \left( {z,\bar z} \right) = X^\mu  \left( z \right) + X^\mu  \left( {\bar z} \right)
\end{equation}
where
\begin{equation}
\begin{gathered}
  X^\mu  \left( z \right) = \frac{{x^\mu  }}
{2} - i\frac{{p^\mu  }}
{{2Q\left( 0 \right)}}\ln \left( z \right) + \frac{i}
{2}\sum\limits_{\omega  \ne 0} {} \frac{{\alpha _\omega ^\mu  }}
{{\omega \sqrt {Q\left( \omega  \right)} }}z^{ - \omega }  \\ 
  X^\mu  \left( {\bar z} \right) = \frac{{x^\mu  }}
{2} - i\frac{{p^\mu  }}
{{2Q\left( 0 \right)}}\ln \left( {\bar z} \right) + \frac{i}
{2}\sum\limits_{\omega  \ne 0} {} \frac{{\alpha _\omega ^\mu  }}
{{\omega \sqrt {Q\left( \omega  \right)} }}\bar z^{ - \omega }  \\ 
\end{gathered} 
\end{equation}
and from this we get
\begin{equation}
\partial X^\mu  \left( z \right) =  - i\frac{{p^\mu  }}
{{2Q\left( 0 \right)z}} - \frac{i}
{2}\sum\limits_{\omega  \ne 0} {} \frac{{\alpha _\omega ^\mu  }}
{{\sqrt {Q\left( \omega  \right)} }}z^{ - \omega  - 1} 
\end{equation}

We will find the singularity at z=w in the OPE of  $\partial X^\mu  \left( z \right)X^\nu  \left( w \right)$
 by considering its vacuum expectation value with $\left| z \right| > \left| w \right|$.

\begin{equation}
\begin{gathered}
  \left\langle 0 \right|\partial X^\mu  \left( z \right)X^\nu  \left( w \right)\left| 0 \right\rangle  = \frac{1}
{4}\sum\limits_{\omega  > 0} {} \left\langle 0 \right|\left( {\frac{{\alpha _\omega ^{^\mu  } }}
{{\sqrt {Q\left( \omega  \right)} }}} \right)\left( {\frac{{\alpha _{ - \omega }^{^\nu  } }}
{{ - \omega \sqrt {Q\left( \omega  \right)} }}} \right)\left| 0 \right\rangle \frac{{w^\omega  }}
{{z^{\omega  + 1} }} \\ 
   =  - \frac{{\eta ^{\mu \nu } }}
{{4z}}\sum\limits_{\omega  > 0} {} \frac{1}
{{m_0 \cos ^2 \left( {\omega \pi } \right) + T_0 \pi }}\left( {\frac{w}
{z}} \right)^\omega   \\ 
\end{gathered} 
\end{equation}
As in Eq.~(\ref{E: Dss}) we can write this sum as a sum of path integrals around allowed frequencies, but here around only the $\omega _n  > 0$ , therefore $n > 0$.
\begin{equation}
\left\langle 0 \right|\partial X^\mu  \left( z \right)X^\nu  \left( w \right)\left| 0 \right\rangle  =  - \frac{{\eta ^{\mu \nu } }}
{{4z}}\sum\limits_{n > 0} {} \oint_{C_n } {\frac{{d\omega }}
{{2\pi i}}} \frac{1}
{{T_0 \cos ^2 \left( {\omega \pi } \right)\left( {\frac{{m_0 \omega }}
{{T_0 }} + \tan \left( {\omega \pi } \right)} \right)}}\left( {\frac{w}
{z}} \right)^\omega  
\end{equation}
As before, this converts to integrals around the zeros of $\cos \left( {\omega \pi } \right)$ , plus integrals along lines parallel to the real axis, which go to zero, plus here an integral along the imaginary $\omega $  axis which however is finite and stays finite as  $z \to w$ .  The result is 
\begin{equation}
\begin{gathered}
  \left\langle 0 \right|\partial X^\mu  \left( z \right)X^\nu  \left( w \right)\left| 0 \right\rangle  =  - \frac{{\eta ^{\mu \nu } }}
{{4\pi T_0 z}}\sum\limits_{k = 0} {} \left( {\frac{w}
{z}} \right)^{\frac{{2k + 1}}
{2}}  \\ 
   =  - \frac{{\eta ^{\mu \nu } }}
{{4\pi T_0 }}\sqrt {\frac{w}
{z}} \frac{1}
{{z - w}} \\ 
   =  - \frac{{\eta ^{\mu \nu } }}
{{4\pi T_0 }}\left( {\frac{1}
{{z - w}} + O\left( {\frac{1}
{w}} \right)} \right) \\ 
\end{gathered} 
\end{equation}

We conclude that
\begin{equation}
\partial X^\mu  \left( z \right)X^\nu  \left( w \right) \sim  - \frac{{\eta ^{\mu \nu } }}
{{4\pi T_0 }}\frac{1}
{{z - w}}
\end{equation}

exactly the same value it would have if there were no attached mass.

\section{7.  Conclusion}

The commutation relations found here are actually what one would  what one would expect if the string position operator was the sum of an ordinary particle operator located at $\sigma  = \pi $  and an independent ordinary open string position operator defined on the interval $0 < \sigma  < \pi$  .

The parameter $m_0 $  measures the mass attached to one end.  The parameter  $T_0 $ measures the mass per unit length.  The results depend greatly on the ratio $r = m_0 /T_0 $  . When r is very large, the frequencies are close to half odd integers, very much what one would expect for a string that is free at one end but fixed at  the other.  

The allowed frequencies $\omega $  are in general not rational numbers.  Therefore although the position operator  $X^\mu  $ can be expressed as a function of the complex variables  $z$ and $\bar z$ ,  and then $X^\mu  \left( {z,\bar z} \right) = X^\mu  \left( z \right) + X^\mu  \left( {\bar z} \right)$, these functions can not be expressed as simple Laurent series in $z$  or $\bar z$  .  However the operator product expansion for two XÕs still has the simple form 

\begin{equation}
X^\mu  \left( z \right)X^\nu  \left( w \right) \sim  - \frac{{\eta ^{\mu \nu } }}
{{4\pi T_0 }}\ln \left( {z - w} \right)
\end{equation}

\end{document}